\documentclass[aoas,MSNbibl,nameyear,dvips]{arximspdf}
\usepackage{graphicx}
\doi{10.1214/13-AOAS668} %kopijuoti is PTS
\volume{7}
\issue{4}
\pubyear{2013}
\firstpage{2272}
\lastpage{2292}

\makeatletter
\newproclaim{remark}{Remark}
\newcommand{\eqref}[1]{(\ref{#1})}
\newcommand\na{n_\mathrm{A}}
\newcommand\nb{n_\mathrm{B}}
\newcommand\bbeta{\bolds{\beta}}
\newcommand\bphi{\bolds{\phi}}
\newcommand\bmeta{\bolds{\eta}}
\newcommand\bGamma{\bolds{\Gamma}}
\newcommand\bLambda{\bolds{\Lambda}}
\newcommand\bSigX{\bolds{\Sigma}_\bX}
\newcommand\bmuX{\bolds{\mu}_\bX}
\newcommand\bX{\mathbf{X}}
\newcommand\bW{\mathbf{W}}
\newcommand\bU{\mathbf{U}}
\newcommand\bIp{\mathbf{I}_p}
\newcommand\var{\operatorname{Var}}
\newcommand\ev{\mathrm{E}}
\newcommand\byA{\mathbf{y}_\mathrm{A}}
\newcommand\bxA{\mathbf{x}_\mathrm{A}}
\newcommand\bwA{\mathbf{w}_\mathrm{A}}
\newcommand\byB{\mathbf{y}_\mathrm{B}}
\newcommand\bxB{\mathbf{x}_\mathrm{B}}
\newcommand\bwB{\mathbf{w}_\mathrm{B}}
\newcommand\trace{\operatorname{Tr}}

\makeatother

\begin{document}
\begin{frontmatter}

\title{Bayesian shrinkage methods for partially observed data with many
predictors\thanksref{T1}}
\runtitle{Bayesian shrinkage}

\thankstext{T1}{Supported by NSF Grant DMS-10-07494 and the National Institutes of Health [CA129102, CA156608, ES020811].}

\begin{aug}
\author[A]{\fnms{Philip S.} \snm{Boonstra}\corref{}\ead[label=e1]{philb@umich.edu}},
\author[A]{\fnms{Bhramar} \snm{Mukherjee}\ead[label=e2]{bhramar@umich.edu}}
\and\break
\author[A]{\fnms{Jeremy M. G.} \snm{Taylor}\ead[label=e3]{jmgt@umich.edu}}
\runauthor{P. S. Boonstra, B. Mukherjee and J. M. G. Taylor}
\affiliation{University of Michigan}
\address[A]{Department of Biostatistics\\
University of Michigan\\
1415 Washington Heights\\
Ann Arbor, Michigan 48109\\
USA\\
\printead{e1}\\
\phantom{E-mail:\ }\printead*{e2}\\
\phantom{E-mail:\ }\printead*{e3}} %adresu isvedimo komanda gale!
\end{aug}

% HISTORY:
\received{\smonth{9} \syear{2012}}
\revised{\smonth{6} \syear{2013}}

% ABSTRACT
%
\begin{abstract}
Motivated by the increasing use of and rapid changes in array
technologies, we consider the prediction problem of fitting a linear
regression relating a continuous outcome $Y$ to a large number of
covariates $\bX$, for example, measurements from current,
state-of-the-art technology. For most of the samples, only the outcome
$Y$ and surrogate covariates, $\bW$, are available. These surrogates
may be data from prior studies using older technologies. Owing to the
dimension of the problem and the large fraction of missing information,
a critical issue is appropriate shrinkage of model parameters for an
optimal bias-variance trade-off. We discuss a variety of fully Bayesian
and Empirical Bayes algorithms which account for uncertainty in the
missing data and adaptively shrink parameter estimates for superior
prediction. These methods are evaluated via a comprehensive simulation
study. In addition, we apply our methods to a lung cancer data set,
predicting survival time ($Y$) using qRT-PCR ($\bX$) and microarray
($\bW$) measurements.
\end{abstract}

% KEYWORDS

\end{frontmatter}

%s1 #&#
\section{Introduction}\label{sec1}
The ongoing development of array technologies for assaying genomic
information has resulted in an abundance of data sets with many
predictors and presents both statistical opportunities and challenges.
As an example, \citet{chen.jto} analyzed a gene-expression microarray
data set of 439 lung adenocarcinomas from four cancer centers in the
United States, with the goal of using gene expression to improve
predictions of survival time relative to using clinical covariates
alone. Expression was measured using Affymetrix oligonucleotide
microarray technology. After pre-screening the probes for consistency
between centers, the authors initially evaluated 13,306 probes for
construction of their predictor.

A clinical challenge to a candidate model which uses Affymetrix data is
its application for predictions in new patients. The underlying
complexity of Affymetrix data, including necessary preprocessing,
requires specialized laboratory facilities, which will be locally
unavailable at smaller hospitals. On the other hand, quantitative
real-time polymerase chain reaction (qRT-PCR) offers a faster and more
efficient assay of the same underlying genomic information, making a
qRT-PCR-based prediction model clinically applicable. The trade-off
comes from the limited number of genes which may be assayed on a single
qRT-PCR card. Thus, from the Affymetrix data, 91 promising genes were
first identified. These 91 genes were then re-assayed with qRT-PCR.
Because of tissue availability issues owing to the multi-center-nature
of the study, only 47 out of 439 tumors were re-assayed by qRT-PCR,
creating a significant missing data problem.

Motivated by this problem, in this paper we consider the analysis of a
data set with many predictors in which a large block of covariates are
missing, a situation for which there is limited previous literature. To
maintain relevance to the application which drives our methodology, we
assume the data have two distinctive features. First, the number of
covariates, that is, genes, is of moderate size, approximately the same
order as the number of observations. This precludes both a more
traditional regression situation as well as an
``ultra-high-dimensional'' regression and reflects that an initial
screening has identified a subset of potentially informative genes.
Second, there are two versions of the genomic data: measurements from a
prior technology, which are complete for all observations, and
measurements from a newer, more efficient technology, which are
observed only on a small subset of the observations. Owing to the
inherent variability in parameter estimates induced by both the missing
data and the dimensionality of the problem, we consider Bayesian
approaches, which allow for the application of shrinkage methods, in
turn offering better prediction.

Translating this into statistical terminology, we consider predicting
an outcome~$Y$ given length-$p$ covariates $\bX$. Assuming $Y$ is
continuous and fully observed, we use the linear model
%
%
%
%e1 #&#
\begin{equation}
Y = \beta_0 + \bX^\top\bbeta+ \sigma\varepsilon.
\label{eqn:linearmodel}
\end{equation}
All observations contain $Y$ and $\bW$, which is an error-prone
length-$p$ surrogate for the true covariate $\bX$. On a small number of
observations of size $\na$, subsample A, we also observe $\bX$, which
is missing for the remaining subjects, constituting subsample B, of
size $\nb$. Complete observations, then, contain an outcome $Y$,
covariates $\bX$ and surrogates $\bW$. Subsample A is written as $\{
\byA, \bxA, \bwA\}$ and subsample B as $\{\byB, \bwB\}$. The true
covariates from subsample B, $\bxB$, are unmeasured. The data are
schematically presented in Figure S1 of the supplemental article
[\citet
{boonstra.aoassupp}].

Our goal is a predictive model for $Y|\bX$ as in equation \eqref
{eqn:linearmodel}, but because $\bW$ is correlated with $\bX$,
subsample B contains information about $\bbeta$. Moreover, shrinkage of
regression coefficients may alleviate problems associated with
multicollinearity of covariates. \citet{boonstra.biostatistics}
proposed a class of targeted ridge (TR) estimators of $\bbeta$,
shrinking estimates toward a target constructed using subsample B,
making a bias-variance trade-off. The amount of shrinkage can be
data-adaptive with a tuning parameter, say, $\lambda$. In a simulation
study of data sets with many predictors, they showed that two biased
methods, a modified regression calibration algorithm and a ``hybrid''
estimator, which is a linear combination of multiple TR estimators with
data-adaptive weights, uniformly out-perform standard regression
calibration, an unbiased method, in terms of mean-squared prediction
error (MSPE):
%
%
%
%e2 #&#
\begin{eqnarray}\label{eqn:mspe}
&&\operatorname{MSPE}(\hat\beta_0,\hat{\bbeta})\nonumber\\
&&\qquad=\ev \bigl[
\bigl(Y_\mathrm{new}-\hat\beta_0-\bX_\mathrm{new}^\top
\hat{\bbeta} \bigr)^2 \bigr]
\\
&&\qquad=\sigma^2+ \bigl(\ev \bigl[\beta_0-\hat
\beta_0+\bX_\mathrm{new}^\top\bbeta-\bX
_\mathrm{new}^\top\hat{\bbeta} \bigr] \bigr)^2+\var
\bigl[\hat\beta_0+\bX_\mathrm{new}^\top\hat{ \bbeta}
\bigr],\nonumber
\end{eqnarray}
where the expectation is over $Y_\mathrm{new},\bX_\mathrm{new},\byA
,\byB
|\bxA,\bwA,\bwB$.

However, there are reasons to consider alternative strategies. The
authors showed the TR estimator can be viewed as a missing data
technique: make an imputation $\tilde{\mathbf{x}}_\mathrm{B}$ of the
missing $\bxB$ and
calculate $\hat{\bbeta}$ treating the data as complete. When the
shrinkage is data-adaptive through the
tuning parameter $\lambda$, there is an intermediate stage: choose
$\lambda$ given $\tilde{\mathbf{x}}_\mathrm{B}$.
Uncertainty in $\tilde{\mathbf{x}}_\mathrm{B}$ or
$\lambda$ is not propagated in the TR estimators, thus, it can be
viewed as improper
imputation [\citet{little.missingdata}]. Moreover, to choose $\lambda$,
a generalized cross-validation (GCV) criterion was applied to subsample
A. Although GCV asymptotically chooses the optimal value of $\lambda$
[\citet{craven.numerische}], it can overfit in finite sample sizes, and
an approach for estimating $\lambda$ which also uses information in
subsample B is preferred. Finally, constructing prediction intervals
corresponding to the point-wise predictions
generated by the class of TR estimators requires use of the
bootstrap. This resampling process is computationally intensive and
provides coverage that may not be nominal.

%
%f1 #&#
\begin{figure}[b]

\includegraphics{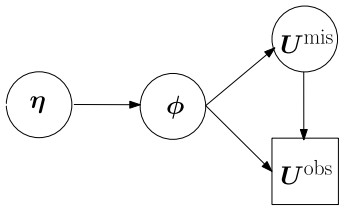}

\caption{A hierarchical model with missing data $\bU
^\mathrm{mis}$ and observed data $\bU^\mathrm{obs}$. The shrinkage penalty
parameters $\bmeta$ are the hyperparameters of $\bphi$, the
quantity(ies) of primary interest.}\label{model:1}
\end{figure}

These reasons, that is, characterizing prediction uncertainty and
unifying shrinkage, imputation of missing data and an adaptive choice
of $\lambda$, motivate a fully Bayesian approach toward the same goal
of improving predictions using auxiliary data. Consider the generic
hierarchical model presented in Figure~\ref{model:1}. Known (unknown,
resp.) quantities are bounded by square (circular) nodes.
Instead of splitting the data into subsamples (cf. Figure S1), we
classify it more broadly into observed $(\bU^\mathrm{obs})$ and missing
$(\bU^\mathrm{mis})$ components. Let $\bphi$ denote parameters of
interest and nuisance parameters in the underlying joint likelihood of
$\{\bU^\mathrm{obs},\bU^\mathrm{mis}\}$. Regularization of $\bphi$ is
achieved through the shrinkage parameter $\bmeta$, equivalently
interpreted in Figure~\ref{model:1} as the hyperparameters which index
a prior distribution on~$\bphi$. One can impose another level of
hierarchy through a hyperprior distribution on~$\bmeta$. Using $[\cdot
]$ and $[\cdot|\cdot]$ to denote
marginal and conditional distributions, draws from $[\bU^\mathrm
{mis},\bphi,\bmeta|\bU^\mathrm{obs}]$, the distribution of unknown random
quantities conditional on the observed data, constitute proper
imputation and incorporate all of the information in the data. Summary
values, like posterior means, as well as measurements of uncertainty,
like highest posterior density credible intervals and prediction
intervals, can easily be calculated based on posterior draws.

Placing the shrinkage parameter $\bmeta$ in a hierarchical framework
allows the flexibility to determine both which components of $\bphi$ to
shrink and to what extent. As an example of the former, \citet
{boonstra.biostatistics} shrink estimates of the regression
coefficients $\bbeta$, tuned by the parameter $\lambda$. However, for
improved prediction of the outcome $Y$, it may be beneficial to shrink
the parameters generating the missing data $\bxB$. For example, in a
nonmissing-data context, the \textsc{scout} method [\citet
{witten.jrssb}] shrinks the estimate of $\var(\bX)$ for better
prediction. As for the extent of shrinkage, the
hyperparameter-equivalence of the tuning parameters allows for the use
of Empirical Bayes algorithms to estimate $\bmeta$. This has been used
in the Bayesian Lasso [\citet{park.jasa,yi.genetics}].

This paper makes two primary contributions. First, in Section~\ref{sec:gibbsvariants} we discuss variants of the Gibbs sampler [\citet
{geman.pami}], a key algorithm for fitting hierarchical models with
missing data. Here, we keep the context broad, assuming a generic
hierarchical model indexed by $\bphi$ with missing data $\bU^\mathrm
{mis}$ and unspecified hyperparameters $\bmeta$, as in Figure~\ref{model:1}. One variant, Data Augmentation [\citet{tanner.jasa}], is a
standard Bayesian approach to missing data, and all unknown quantities
have prior distributions. Two others are Empirical Bayes methods: the
Monte Carlo expectation--maximization algorithm [\citet{wei.jasa}] and
the Empirical Bayes Gibbs sampler [\citet{casella.biostatistics}].
Although proposed for seemingly different problems, we argue that the
sampling strategies in each are special cases of that in Figure~\ref{model:1}: variants of the same general algorithm, which we call
EM-within-Gibbs. This previously-unrecognized link is important, given
the increasing role Empirical Bayes methods play in modern
applications. The second primary contribution builds on this proposed
framework (Section~\ref{sec:likelihoodspec}), namely, a comparison of
several fully Bayesian and Empirical Bayes options and their
application to our motivating genomic analysis. Of note in the data are
two crucial features: first, $\bphi$, comprised of $\beta_0$, $\bbeta$,
$\sigma^2$ plus parameters for modeling the distribution of $\bX$, is
of a significant dimension, so that fitting a model with \emph{no}
missing data would still be somewhat challenging, and, second, the
number of partial observations where $\bX$ is missing is larger than
the number of complete observations. Meaningful analysis then requires
the regularization, or shrinkage, of $\bphi$ via an appropriate
specification of the hierarchy and choice of $\bmeta$. We propose to
shrink several different components of $\bphi$, making use of the
simultaneous interpretation of $\bmeta$ as a shrinkage penalty and a
hyperparameter on $\bphi$. We evaluate these methods via a
comprehensive simulation study (Section~\ref{sec:simulations}), also
considering robustness of these methods under model misspecification.
Finally, we analyze the \citeauthor{chen.jto} data set (Section~\ref{sec:dataanalysis}). We include ridge regression [\citet
{hoerl.technometrics}] as a reference, because the additional modeling
assumptions of the other likelihood-based methods offer efficiency
gains only when they are satisfied.

%s2 #&#
\section{Gibbs sampler variants}
\label{sec:gibbsvariants}
In this section we discuss four existing variants of the Gibbs sampler
relevant to our analysis. As we will argue, two of these are special
cases of a more general variant, which we call ``Empirical Bayes Within
Gibbs'' (\textbf{EWiG}), an equivalence that has not been established
previously, leaving three distinct variants. We define a ``variant''
here as the characterization of a posterior distribution plus an
algorithm for fitting the model. All variants are summarized in Table~\ref{tab:1}.

%t1 #&#
\begin{table}
\tabcolsep=1pt
\caption{A comparison of the general form of the Gibbs sampler variants
from Section \protect\ref{sec:gibbsvariants} as they were originally proposed.
Differences between posteriors depend on the presence of missing data
$\bU^\mathrm{mis}$ and whether the hyperparameters $\bmeta$ are fully
known. Differences in algorithms depend on how the lowest level of the
hierarchy, which is unknown, is treated. In particular, \textbf{MCEM}
differs from \textbf{DA} because it returns only an estimate of the
posterior mode}\label{tab:1}
\begin{tabular*}{\textwidth}{@{\extracolsep{\fill}}lcc@{}}
\hline
\textbf{Variant} & \textbf{Posterior} & \textbf{Prior on} $\bolds{\bmeta}$\\
\hline
\textbf{DA} [\citet{tanner.jasa}]& $[\bphi,\bU^\mathrm{mis}|\bU
^\mathrm
{obs},\bmeta]\propto[\bU^\mathrm{obs},\bU^\mathrm{mis}|\bphi
]\times[\bphi
|\bmeta]$ & No \\
\textbf{DA+} [\citet{gelfand.jasa}]& $[\bphi,\bU^\mathrm
{mis},\bmeta|\bU
^\mathrm{obs}]\propto[\bU^\mathrm{obs},\bU^\mathrm{mis}|\bphi
]\times[\bphi
|\bmeta]\times[\bmeta]$ & Yes \\
\textbf{MCEM} [\citet{wei.jasa}]& $[\bphi,\bU^\mathrm{mis}|\bU
^\mathrm
{obs},\bmeta]\propto[\bU^\mathrm{obs},\bU^\mathrm{mis}|\bphi
]\times[\bphi
|\bmeta]$ & No \\
\textbf{EBGS} [\citet{casella.biostatistics}] & $[\bphi|\bU^\mathrm
{obs},\bmeta]\propto[\bU^\mathrm{obs}|\bphi]\times[\bphi|\bmeta
]$&No\\
\textbf{EWiG} & $[\bphi,\bU^\mathrm{mis}|\bU^\mathrm{obs},\bmeta
]\propto[\bU
^\mathrm{obs},\bU^\mathrm{mis}|\bphi]\times[\bphi|\bmeta]$&No\\
\hline
\end{tabular*}
\end{table}

\textbf{Data augmentation (DA+, DA)} [\citet{tanner.jasa}].
\begin{eqnarray*}
&&\mathit{Posterior}\textup{:}\ \bigl[\bphi,\bU^\mathrm{mis},\bmeta|
\bU^\mathrm{obs} \bigr]\propto \bigl[\bU^\mathrm{obs},
\bU^\mathrm{mis}|\bphi \bigr]\times[\bphi|\bmeta]\times[\bmeta]
\\
&&\mathit{Algorithm}\textup{:}\ \mbox{ at iteration } i,
\nonumber
\\
&&\phantom{\textit{Algorithm}: }\quad \bU^\mathrm{mis} {}^{(i)}\sim \bigl[\bU^\mathrm{mis}|
\bU^\mathrm{obs},\bphi^{(i-1)} \bigr]
\\
&&\phantom{\textit{Algorithm}: } \quad\bphi^{(i)} \sim \bigl[\bphi|\bU^\mathrm{obs},
\bU^\mathrm {mis} {}^{(i)},\bmeta^{(i-1)} \bigr]
\\
&&\phantom{\textit{Algorithm}: }\quad \bmeta^{(i)}\sim \bigl[\bmeta|\bphi^{(i)} \bigr].
\end{eqnarray*}
These two variants are natural Bayesian treatments of missing data:
$\bU
^\mathrm{mis}$ and $\bphi$ are both unobserved random variables. In
\textbf{DA+}, which is given above, the hyperparameters $\bmeta$ are
also unknown [\citet{gelfand.jasa}]. In \textbf{DA}, a value for
$\bmeta
$ is chosen. In either case, draws of $\bphi$ and $\bU^\mathrm{mis}$ are
sequentially made from their conditional posteriors. In \textbf{DA+}
only, $\bmeta$ is also sampled from its conditional posterior. Then, in
either \textbf{DA} or \textbf{DA+}, the whole process is iterated.
Tanner and Wong prove that iterations will eventually yield a draw from
the true posterior distribution of interest, $[\bphi,\bU^\mathrm
{mis},\bmeta|\bU^\mathrm{obs}]$ for \textbf{DA+} or $[\bphi,\bU
^\mathrm
{mis}|\bU^\mathrm{obs},\bmeta]$ for \textbf{DA}. The full conditional
distribution $[\bphi|\bU^\mathrm{obs},\bU^\mathrm{mis},\bmeta]$
may be
difficult to specify. Suppose instead a set of partial conditional
distributions is available, $[\bphi_{J}|\bphi_{(J)},\bU^\mathrm
{obs},\bU
^\mathrm{mis},\bmeta]$, where the set of $J$'s forms a partition of the
vector $\bphi$. Then under mild conditions, repeated iterative sampling
from these partial conditional distributions will also yield draws from
the true posterior [\citet{geman.pami}].

\textbf{Monte Carlo expectation--maximization (MCEM)}
[\citet
{wei.jasa}].
\begin{eqnarray*}
&&\textit{Posterior}\textup{:}\  \bigl[\bphi,\bU^\mathrm{mis}|\bU^\mathrm
{obs},\bmeta \bigr]\propto \bigl[\bU^\mathrm{obs},\bU^\mathrm{mis}|
\bphi \bigr]\times[\bphi|\bmeta]
\\
&&\textit{Algorithm}\textup{:}\ \mbox{at iteration } i,
\nonumber
\\
&&\phantom{\textit{Algorithm: }}\quad \mbox{for } k=1,\ldots,K,
\nonumber
\\
&&\phantom{\textit{Algorithm: }}\qquad \bU^\mathrm{mis} {}^{(i,k)}\sim \bigl[\bU^\mathrm{mis}|
\bU^\mathrm{obs},\bphi^{(i-1)} \bigr]
\\
&&\phantom{\textit{Algorithm: }}\qquad \bphi^{(i)} = \mathop{\arg\max}_{\bphi}\dfrac{1}{K} \sum
_{k=1}^K \ln \bigl[\bphi|
\bU^\mathrm{obs},\bU^\mathrm{mis} {}^{(i,k)},\bmeta \bigr].
\end{eqnarray*}
\textbf{MCEM} provides a point estimate of $\bphi$ rather than an
estimate of the posterior distribution, as with \textbf{DA}/\textbf
{DA+}. It is a modification of the original EM algorithm [\citet
{dempster.jrssb}], replacing an intractable expectation with a Monte
Carlo average of multiple imputations. $K$ draws of $\bU^\mathrm{mis}$
are sampled conditional on the current value of $\bphi\dvtx \bphi
^{(i-1)}$. The expected posterior is updated with a Monte Carlo average
and maximized with respect to $\bphi$. When $\bphi$ has a flat prior,
as in the originally proposed \textbf{MCEM}, $\{\bphi^{(i)}\}$ will
converge to the maximum likelihood estimate (MLE) of $\bphi$. If an
informative prior is specified through a particular choice of $\bmeta$,
the sequence will converge to a \emph{penalized} MLE [\citet{green.jrssb}].

\textbf{Empirical Bayes Gibbs sampling (EBGS)} [\citet
{casella.biostatistics}].
\begin{eqnarray*}
&&\textit{Posterior}\textup{:}\  \bigl[\bphi|\bU^\mathrm{obs},\bmeta \bigr]\propto
\bigl[\bU^\mathrm{obs}|\bphi \bigr]\times[\bphi|\bmeta]
\\
&&\textit{Algorithm}\textup{:}\ \mbox{at iteration } i,
\\
&&\phantom{\textit{Algorithm: }} \quad\mbox{for } k=1,\ldots,K,
\nonumber
\\
&&\phantom{\textit{Algorithm: }}\qquad \bphi^{(i,k)}\sim \bigl[\bphi|\bU^\mathrm{obs},
\bmeta^{(i-1)} \bigr]
\\
&&\phantom{\textit{Algorithm: }} \qquad\bmeta^{(i)} = \mathop{\arg\max}_{\bphi}\dfrac{1}{K} \sum
_{k=1}^K \ln \bigl[\bphi^{(i,k)}|
\bmeta \bigr].
\end{eqnarray*}
\textbf{EBGS} allows the data to determine a value for the
hyperparameter $\bmeta$. In the context of Casella, there are no
missing data $\bU^\mathrm{mis}$. However, $\bphi$ is considered missing
for purposes of determining $\bmeta$: choose $\bmeta$ which maximizes
its \emph{marginal} log-likelihood, $\ln[\bU^\mathrm{obs}|\bmeta
]$. As
in \textbf{MCEM}, an EM-type algorithm can maximize this intractable
log-likelihood. $K$ draws of $\bphi$ are made from the current estimate
of its posterior, and $\bmeta$ is updated by maximizing a Monte Carlo
estimate of $\ev[\ln[\bphi|\bmeta] ]$, where the expectation
is over the distribution $[\bphi|\bU^\mathrm{obs},\bmeta^{(i)}]$. This
expected complete-data log-likelihood relates to the desired marginal
log-likelihood as follows. First observe that
\begin{eqnarray*}
{ \bigl[\bU^\mathrm{obs}|\bmeta \bigr] } { \bigl[\bphi|
\bU^\mathrm{obs},\bmeta \bigr] }&=&{[\bphi|\bmeta] } { \bigl[
\bU^\mathrm{obs}|\bphi,\bmeta \bigr]}
\\
&=&[\bphi|\bmeta] \bigl[\bU^\mathrm{obs}|\bphi \bigr].
\end{eqnarray*}
Let $C=\ev[\ln[\bU^\mathrm{obs}|\bphi] ]$, which is constant
with respect to $\bmeta$. Then,
\[
\ln \bigl[\bU^\mathrm{obs}|\bmeta \bigr] = \ev \bigl[ \ln[\bphi |\bmeta]
\bigr]- \ev \bigl[ \ln \bigl[\bphi|\bU^\mathrm{obs},\bmeta \bigr] \bigr]+
C.
\]
Because $\ev[ \ln[\bphi|\bU^\mathrm{obs},\bmeta]
]\leq\ev
[ \ln[\bphi|\bU^\mathrm{obs},\bmeta^{(i)}] ]$ for any
$\bmeta
$, we have the result that maximizing $ \ev[ \ln[\bphi|\bmeta]
]$ (or a Monte Carlo approximation thereof) over $\bmeta$ will
increase $ \ln[\bU^\mathrm{obs}|\bmeta] $ and converge to a local maximum.

\textbf{EM-within-Gibbs (EWiG)}.
\begin{eqnarray*}
&&\textit{Posterior}\textup{:}\  \bigl[\bphi,\bU^\mathrm{mis}|\bU^\mathrm
{obs},\bmeta \bigr]\propto \bigl[\bU^\mathrm{obs},\bU^\mathrm{mis}|
\bphi \bigr]\times[\bphi|\bmeta]
\\
&&\textit{Algorithm}\textup{:}\ \mbox{at iteration } i,
\\
&&\phantom{\textit{Algorithm: }} \quad\mbox{for } k=1,\ldots,K,
\\
&&\phantom{\textit{Algorithm: }}\qquad \bU^\mathrm{mis} {}^{(i,k)}\sim \bigl[\bU^\mathrm{mis}|
\bU^\mathrm{obs},\bphi^{(i,k-1)} \bigr]
\\
&&\phantom{\textit{Algorithm: }}\qquad \bphi^{(i,k)}\sim \bigl[\bphi|\bU^\mathrm{obs},\bU^\mathrm
{mis} {}^{(i,k)},\bmeta^{(i-1)} \bigr]
\\
&&\phantom{\textit{Algorithm: }} \qquad\bmeta^{(i)} = \mathop{\arg\max}_{\bphi}\dfrac{1}{K} \sum
_{k=1}^K \ln \bigl[\bphi^{(i,k)}|
\bmeta \bigr].
\end{eqnarray*}
Importantly, both \textbf{MCEM} and \textbf{EBGS} allow the lowest
level of the hierarchy to be adaptively determined by the data rather
than chosen a priori. In \textbf{MCEM}, this lowest level is $\bphi$,
and in \textbf{EBGS}, it is $\bmeta$. However, \textbf{MCEM} can be
expanded in the presence of an unknown $\bmeta$ by putting both $\bU
^\mathrm{mis}$ and $\bphi$ into the imputation step, so $\bphi$ is
sampled rather than optimized. The maximization step determines $\bmeta
$. This returns to the original goal of \textbf{DA+}/\textbf{DA}, which
is determining the posterior distribution of $\bphi$. Equivalently, we
can take the perspective of expanding \textbf{EBGS}: add an imputation
step for sampling $\bU^\mathrm{mis}$, keeping the maximization step the
same. As a result of this equivalence, expanding either \textbf{MCEM}
or \textbf{EBGS} yields the same result, what we call \textbf{EWiG},
given above. Because $\bmeta$ is unknown, the hierarchical model here
is the same as that given in Figure~\ref{model:1}.

In summary, we have asserted that \textbf{MCEM} and \textbf{EBGS} are
special cases of \textbf{EWiG}, so there are three distinct variants
which we apply to our problem in the following section: \textbf{DA},
\textbf{DA+}, and \textbf{EWiG}.

%

%s3 #&#
\section{Specification of likelihood and priors}
\label{sec:likelihoodspec}
The discussion so far has been deliberately generic. We now specify a
likelihood for our problem of interest, which in turn gives $\bphi$,
and apply these Gibbs variants to several combinations of (i) choices
of priors $[\bphi|\bmeta]$ and (ii) values of the hyperparameter
$\bmeta$.
Translating the quantities in Figure~\ref{model:1} to our problem, we have
$\bU^\mathrm{obs}=\{\byA,\byB,\bxA,\bwA,\bwB\}$ and
$\bU^\mathrm{mis}=\bxB$. A commonly used factorization of the joint
likelihood is
$[Y,\bX,\bW]=[Y|\bX][\bW|\bX][\bX]$, which makes a conditional
independence assumption $[Y|\bX,\bW]=[Y|\bX]$. An alternative
factorization is $[Y|\bX][\bX|\bW]$, which we do not consider, as it is
inconsistent with the application-driven measurement error structure of
$\bW$ and $\bX$. We make the following assumptions:
%
%
%
%e3 #&#
\begin{eqnarray}\label{eqn:models}
Y|\bX&=& N \bigl\{\beta_0 + \bX^\top\bbeta,
\sigma^2 \bigr\},\qquad \bW|\bX= N_p \bigl\{ \psi
\mathbf{1}_p + \nu\bX, \tau^2\bIp \bigr\},
\nonumber
\\[-8pt]
\\[-8pt]
\nonumber
 \bX&=&
N_p \{\bmuX,\bSigX\}.
\end{eqnarray}
The likelihood has an outcome model relating $Y$ to $\bX$, a
measurement error
model relating the error-prone $\bW$ to $\bX$, and a multivariate
distribution for $\bX$. Thus, $\bphi=\{\beta_0,\bbeta,\sigma,\psi
,\nu
,\tau,\bmuX,\bSigX\}$, and $\bmeta$ is described below. Of interest is
prediction of a new value
$Y_\mathrm{new}$ given $\bX_\mathrm{new}$, for example, $\hat
Y_\mathrm{new}=\beta_0^*+\bX_\mathrm{new}^\top\bbeta^*$, where
$\beta_0^*$ and $\bbeta^*$ are posterior summaries of $\beta_0$ and
$\bbeta$. Uncertainty is quantified using the empirical distribution of
$\hat
Y_\mathrm{new}^{(t)}=\beta_0^{(t)}+\bX_\mathrm{new}^\top\bbeta
^{(t)} +
\sigma^2{}^{(t)}\varepsilon^{(t)}$, where
$\{\beta_0^{(t)},\bbeta^{(t)},\sigma^2{}^{(t)}\}$ is the set of
posterior draws and
$\varepsilon^{(t)}\stackrel{\mathrm{i.i.d.}}{\sim}N\{0,1\}$. If $\bxB
=\bU
^\mathrm{mis}$ were observed, the \emph{complete} log-likelihood
would be
%
%
%
%e4 #&#
\begin{eqnarray}
\label{eqn:LC} \ell_C&=&\ln \bigl[\bU^\mathrm{obs},
\bU^\mathrm{mis}| \bphi \bigr]\nonumber\\
&=&\ln \bigl[\byA|\bxA,\beta_0,
\bbeta,\sigma^2 \bigr]+\ln \bigl[\bwA|\bxA,\psi,\nu,\tau^2
\bigr]+\ln[\bxA |\bmuX, \bSigX]
\\
&&{} +\ln \bigl[\byB|\bxB,\beta_0,\bbeta,\sigma^2 \bigr]+
\ln \bigl[\bwB|\bxB,\psi,\nu,\tau^2
\bigr]+\ln[\bxB|\bmuX,\bSigX].\nonumber
\end{eqnarray}
The log-likelihood gives the imputation step:
%
%
%
%e5 #&#
\begin{equation}
\bxB|\bU^\mathrm{obs},\bphi= N_{\nb\times p} \bigl\{\tilde{\mathbf{x}}_\mathrm{B}  \bigl(
\bU^\mathrm{obs},\bphi \bigr),\bGamma(\bphi) \bigr\}\label{eqn:xbimp},
\end{equation}
where $\bGamma(\bphi)=[\bbeta\bbeta^\top/\sigma^2+(\nu^2/\tau
^2)\bIp
+{\bolds{\Sigma}_\bX^{-1}}]^{-1}$ and $\tilde{\mathbf{x}}_\mathrm{B}(\bU^\mathrm
{obs},\bphi
)=[(\byB-\break \beta
_0\mathbf{1}_{\nb})\bbeta^\top/\sigma^2+(\nu/\tau^2)(\bwB-\psi\mathbf{1}_{\nb}\mathbf{1}_p^\top)+(\mathbf{1}_{n_\mathrm{B}}\bmuX^\top){\bolds{\Sigma}_\bX
^{-1}}]\bGamma(\bphi)$.
Note that the mean is an $\nb\times p$ matrix, each row representing
the mean vector corresponding to a length-$p$ observation, but the
covariance is shared. The imputation is defined only by the likelihood
and is common to all methods we consider; the differences lie in the
choice of prior $[\bphi|\bmeta]$ and the hyperparameter $\bmeta$. These
crucially determine the nature and extent of shrinkage induced on
$\bphi
$. In what follows, we propose several options, summarized in Table~\ref{tab:2}.

%t2 #&#
\begin{table}
\tabcolsep=0pt
\caption{A summary of all Gibbs samplers and choices of priors we
considered. $\bolds{\Lambda}$ is constrained to the class of diagonal
matrices. \textsc{vanilla} and \textsc{ebsigmax} require that $p\leq
\na+\nb$}\label{tab:2}
\begin{tabular*}{\textwidth}{@{\extracolsep{\fill}}lcccc@{}}
\hline
\textbf{Method} & $\bolds{[\bbeta|\bmeta]\propto}$ & $\bolds{[{\bolds{\Sigma}_\bX
^{-1}}|\bmeta]\propto
|\bSigX
^{-1}|^{(2p-1)/2}\times}$ &$\bolds{\bmeta}$&\textbf{Variant} \\
\hline
\textsc{vanilla}&1&$\exp\{-\frac{2p-1}{2}\trace(\operatorname
{diag}(\widehat{\var}[\bxA])\bSigX^{-1}) \}$&$\{\}$&\textbf{DA}\\[3pt]
\textsc{hierbetas}&$ (\frac{\lambda}{\sigma^2}
)^{p/2}\exp
\{-\frac{1}{2}\frac{\lambda}{\sigma^2}\bbeta^\top\bbeta\}
$&$\exp
\{-\frac{2p-1}{2}\trace(\operatorname{diag}(\widehat{\var}[\bxA
])\bSigX
^{-1}) \}$&$\{\lambda\}$&\textbf{DA+}\\[3pt]
\textsc{ebbetas}&$ (\frac{\lambda}{\sigma^2} )^{p/2}\exp
\{
-\frac{1}{2}\frac{\lambda}{\sigma^2}\bbeta^\top\bbeta\}
$&$\exp
\{-\frac{2p-1}{2}\trace(\operatorname{diag}(\widehat{\var}[\bxA
])\bSigX
^{-1}) \}$&$\{\lambda\}$&\textbf{EWiG}\\[3pt]
\textsc{ebsigmax}&1&$|\bolds{\Lambda}|^{3p/2}\exp\{-(1/2)\trace
(\bm\Lambda\bSigX^{-1}) \}$&$\{\bolds{\Lambda}\}$&\textbf{EWiG}\\[3pt]
\textsc{ebboth}&$ (\frac{\lambda}{\sigma^2} )^{p/2}\exp
\{
-\frac{1}{2}\frac{\lambda}{\sigma^2}\bbeta^\top\bbeta\}
$&$|\bm\Lambda|^{3p/2}\exp\{-(1/2)\trace(\bolds{\Lambda}\bSigX
^{-1}) \}
$&$\{\lambda,\bolds{\Lambda}\}$&\textbf{EWiG}\\
\hline
\end{tabular*}  \vspace*{-3pt}
\end{table}

{\textsc{vanilla}}. As a baseline approach, we apply \textbf{DA}
to the problem. The choice of prior is
%
%
%
%e6 #&#
\begin{equation}
\qquad[\bphi|\bmeta]\propto \bigl(\sigma^2\tau^2
\bigr)^{-1}\bigl|\bSigX^{-1}\bigr|^{(2p-1)/2}\exp \biggl\{-
\frac{2p-1}{2}\trace \bigl(\operatorname{diag} \bigl(\widehat{\var }[\bxA] \bigr)
\bSigX ^{-1} \bigr) \biggr\}\label{eqn:flatprior},
\end{equation}
where $\operatorname{diag}(\widehat{\var}[\bxA])$ is the diagonal part
of the
empirical covariance of $\bxA$. This is a Jeffreys prior on each
component of $\bphi$ except ${\bolds{\Sigma}_\bX^{-1}}$ (see Remark
\ref{rem1} below), and
$\bmeta$ is known. The product of expressions \eqref{eqn:LC} and
\eqref
{eqn:flatprior} yields the full conditional distributions of each
component of $\bphi$. For brevity, we present only the Gibbs steps for
$\bbeta$ and ${\bolds{\Sigma}_\bX^{-1}}$; the complete set of full
conditional
distributions are given in the supplemental article [\citet
{boonstra.aoassupp}]:
%
%
%
%e7 #&#
\begin{eqnarray}\label {eqn:gibbsbetas}
\bbeta&\sim& N_p \bigl\{ \bigl(\bxA^\top\bxA+
\bxB^\top\bxB \bigr)^{-1} \bigl(\bxA^\top[\byA-
\beta_0\mathbf{1}_{\na}]+\bxB^\top[\byB-
\beta_0\mathbf{1}_{\nb
}] \bigr),\nonumber\\
&&\hspace*{170pt}{}\sigma^2 \bigl(
\bxA^\top\bxA+ \bxB^\top\bxB \bigr)^{-1} \bigr\},
\nonumber\\
{\bolds{\Sigma}_\bX^{-1}}&\sim& W \bigl\{3p+\na+\nb,\\
&&\hspace*{16pt}{}
\bigl((2p-1)\operatorname{diag} \bigl(\widehat{\var}[\bxA] \bigr) + \bigl(\bxA- \mathbf{1}_{\na}\bmuX^\top \bigr)^\top
\bigl(\bxA- \mathbf{1}_{\na}\bmuX^\top \bigr)\nonumber \\
&&\hspace*{107pt}{}+ \bigl(\bxB-
\mathbf{1}_{\nb}\bmuX^\top \bigr)^\top \bigl(\bxB-
\mathbf{1}_{\nb}\bmuX^\top \bigr) \bigr)^{-1} \bigr\}.\nonumber
\end{eqnarray}
The Wishart distribution with $d$ degrees of freedom, $W\{d,\mathbf{S}\}$,
has mean $d\mathbf{S}$. As made clear in the matrix inversion in \eqref
{eqn:gibbsbetas}, \textsc{vanilla} may only be implemented when $p\leq
\na+\nb$.

%
%re1 #&#
\begin{remark}\label{rem1}
A Jeffreys prior on ${\bolds{\Sigma}_\bX^{-1}}$, ${\bolds{\Sigma
}_\bX^{-1}}\sim W\{0,0\bIp\}$,
may result in an improper joint posterior if $\nb\gg\na$ and $p$ is
large, that is, when the fraction of missing data is large. From our
numerical studies and monitoring of trace plots, even a minimally
proper prior on ${\bolds{\Sigma}_\bX^{-1}}$, that is, using $p+1$
degrees of freedom,
does not ensure a proper posterior. We assume a priori ${\bolds{\Sigma
}_\bX^{-1}}\sim
W \{3p,(2p-1)^{-1}[\operatorname{diag}(\widehat{\var}[\bxA])]^{-1} \}
$, a
data-driven choice, the density of which is given in~\eqref
{eqn:flatprior}. The prior mean of ${\bolds{\Sigma}_\bX^{-1}}$ is
$\frac
{3p}{2p-1}[\operatorname{diag}(\widehat{\var}[\bxA])]^{-1}$, and the
prior mean of
$\bSigX$ is $\operatorname{diag}(\widehat{\var}[\bxA])$. Heuristic
numeric evidence
shows that $3p$ degrees of freedom works well, but we have not
demonstrated a theoretical optimality for this. Other values that
ensure convergence are equally defensible.
\end{remark}

We call the Gibbs sampler which uses this mildly informative prior
specification \textsc{vanilla}. All the other methods we propose will
have modified Gibbs steps for two components of $\bphi\dvtx \bbeta$ and
${\bolds{\Sigma}_\bX^{-1}}$. Shrinking $\bbeta$ is a clear choice:
from \eqref
{eqn:models}, $\bbeta$ is closely tied to prediction of $Y|\bX$. As for
${\bolds{\Sigma}_\bX^{-1}}$, this determines in part the posterior
variance of $\bxB$
\eqref{eqn:xbimp}; as this variance increases, the posterior variance
of $\bbeta$ decreases~\eqref{eqn:gibbsbetas}, thereby shrinking draws
$\bbeta$. Other factors in the variance of $\bxB$, like~$\tau^2$, are
additional candidates for shrinkage, but we do not pursue this here.

%s3.1 #&#
\subsection{\texorpdfstring{Adaptive prior on $\beta$}{Adaptive prior on beta}}%{Adaptive prior on $\bolds{\beta}$}

Since we are interested in regularizing predictions of the outcome $Y$,
a natural candidate for shrinkage via an informative prior is the
parameter vector $\bbeta$, which yields the conditional mean of $Y|\bX
$. Ridge regression offers favorable predictive capabilities [\citet
{frank.technometrics}], and the $\ell_2$ penalty on the norm of
$\bbeta
$ is conjugate to the Normal log-likelihood. For these reasons, we
replace the Jeffreys prior on $\bbeta$ in \eqref{eqn:flatprior} with
%
%
%
%e8 #&#
\begin{equation}
\bigl[\bbeta|\sigma^2,\lambda \bigr]\propto \biggl(\frac{\lambda
}{\sigma
^2}
\biggr)^{p/2}\exp \biggl\{-\frac{1}{2}\frac{\lambda}{\sigma^2}
\bbeta^\top\bbeta \biggr\}.\label{eqn:ridgeprior}
\end{equation}
This normal prior on $\bbeta$ is analogous to Bayesian ridge
regression. $\lambda$ is a hyperparameter, that is, $\bmeta=\{\lambda
\}$.
Conditional upon $\lambda$, the Gibbs step for $\bbeta$ is\looseness=-1
\begin{eqnarray*}
\bbeta&\sim& N_p \bigl\{ \bigl(\bxA^\top\bxA+ {\mathbf{x}}{}^\top_\mathrm{B} {\mathbf{x}}_\mathrm{B}+\lambda
\mathbf{I}_p \bigr)^{-1} \bigl(\bxA^\top\byA+{\mathbf{x}}{}^\top_\mathrm{B} \byB \bigr),\\
&&\hspace*{85pt}{}\sigma^{2} \bigl(
\bxA^\top\bxA+ {\mathbf{x}}{}^\top_\mathrm{B} {\mathbf{x}}_\mathrm{B}+\lambda\mathbf{I}_p \bigr)^{-1} \bigr\}.
\end{eqnarray*}\looseness=0
Thus, the posterior mean of $\bbeta$ is shrunk toward zero and with
smaller posterior variance. As we have outlined in Section~\ref{sec:gibbsvariants}, there are several options for the treatment of
$\lambda$.\vadjust{\goodbreak}

{\textsc{hierbetas}}. Following \citet{gelfand.jasa}, we can
treat the hyperparameter $\lambda$ as random (\textbf{DA+}) with prior
distribution $[\lambda]\propto\lambda^{-1}$. Then, we have the
following additional posterior step: $\lambda\sim G \{p/2,\bbeta
^\top\bbeta/(2\sigma^2) \}$. This Bayesian ridge regression with
posterior sampling of $\lambda$ is denoted by \textsc{hierbetas}.

{\textsc{ebbetas}}. Alternatively, we may apply \textbf
{EWiG} to
estimate $\lambda$. That is, integrate $\log[\bbeta|\sigma
^2,\lambda]$
with respect to the density $[\bphi|\bU^\mathrm{obs},\lambda]$,
differentiate with respect to $\lambda$, and solve for $\lambda$. The
resulting \textbf{EWiG} update is
 $\lambda\leftarrow p [(1/K)\sum_{k=1}^K \bbeta^{(k)}{}^\top\bbeta^{(k)}{}/\sigma^{2(k)} ]^{-1}$.
This is a Monte Carlo estimate of\break  $p \{\ev[(\bbeta^\top
\bbeta
)/(\sigma^2) ] \}^{-1}$, the maximum of the marginal
likelihood of $\lambda$. The update occurs at every $K$th iteration of
the algorithm using the previous $K$ draws of $\bbeta$ and $\sigma^2$;
larger values of $K$ yield a more precise estimate. This Bayesian ridge
with an Empirical Bayes update of $\lambda$ is denoted by \textsc{ebbetas}.

%s3.2 #&#
\subsection{\texorpdfstring{Adaptive prior on $\Sigma_\bX^{-1}$ ({\textsc{ebsigmax}}, {\textsc{ebboth}})}
{Adaptive prior on Sigma X^{-1} ({\textsc{ebsigmax}}, {\textsc{ebboth}})}}
We noted previously that an
informative prior on ${\bolds{\Sigma}_\bX^{-1}}$ is necessary to
ensure a proper joint
posterior: ${\bolds{\Sigma}_\bX^{-1}}\sim W \{
3p,(2p-1)^{-1}[\operatorname{diag}(\widehat{\var}
[\bxA])]^{-1} \}$, which has inverse scale matrix $(2p-1)\operatorname
{diag}(\widehat{\var}[\bxA])$. As we have noted, shrinkage of ${\bolds
{\Sigma}_\bX^{-1}}$ is
closely related to that of $\bbeta$. This was exploited by \citet
{witten.jrssb} in the \textsc{scout} procedure, suggesting that
prediction can be improved through adaptive regularization of
${\bolds{\Sigma}_\bX^{-1}}
$. Leaving the inverse scale matrix unspecified, the prior is
%{}
%
%
%e9 #&#
\begin{equation}
\bigl[{\bolds{\Sigma}_\bX^{-1}}|\bolds{\Lambda} \bigr]\propto|\bm\Lambda |^{3p/2}\bigl|\bSigX ^{-1}\bigr|^{(2p-1)/2}\exp \bigl\{-(1/2)
\trace \bigl(\bolds{\Lambda}\bSigX ^{-1} \bigr) \bigr\} .\label{eqn:sigxprior}
\end{equation}
$\bolds{\Lambda}$ is the unknown positive-definite matrix of
hyperparameters. The full conditional distribution of ${\bolds{\Sigma
}_\bX^{-1}}$ becomes
%
%
%
%e10 #&#
\begin{eqnarray}\label{eqn:sigxconddist}
&&{\bolds{\Sigma}_\bX^{-1}}\sim W \bigl\{3p+n_\mathrm{A}+n_\mathrm{B},\nonumber\\
&&\hspace*{26pt}\qquad \bigl(\bolds{\Lambda}+ \bigl(\bxA- \mathbf{1}_{\na}\bmuX^\top
\bigr)^\top \bigl(\bxA- \mathbf{1}_{\na}\bmuX^\top \bigr) \\
&&\hspace*{29pt}\qquad\quad{}+
\bigl(\bxB- \mathbf{1}_{\nb}\bmuX^\top \bigr)^\top \bigl(
\bxB- \mathbf{1}_{\nb
}\bmuX^\top \bigr) \bigr)^{-1} \bigr\}
 .\nonumber
\end{eqnarray}
$\bolds{\Lambda}$ may be random or it can be updated with an \textbf{EWiG}
step. Given the potential difficulty in precisely estimating an
unconstrained matrix which maximizes the marginal likelihood, we
constrain $\bolds{\Lambda}$ to be diagonal. Under this constraint, the
\textbf{EWiG} update for the $i$th diagonal of $\bolds{\Lambda}$ is
$\Lambda
_{ii}\leftarrow3p ((1/K)\sum_{k=1}^K{\bolds{\Sigma}_\bX^{-1}}
_{(ii)}^{(k)}
)^{-1}$, where ${\bolds{\Sigma}_\bX^{-1}}_{(ii)}$ indicates the
$i$th diagonal element
of ${\bolds{\Sigma}_\bX^{-1}}$. Then, $\bolds{\Lambda}= \operatorname
{diag}\{\Lambda
_{11},\ldots
, \Lambda_{pp}\}$. This is a Monte Carlo approximation of
$3p\operatorname
{diag}\{\ev[\bSigX^{-1}]_{11},\ldots, \ev[\bSigX^{-1}]_{pp}\}
^{-1}$, the
minimizer of $\ev[\log[{\bolds{\Sigma}_\bX^{-1}}|\bolds{\Lambda}] ]$
with respect
to $\bolds{\Lambda}$, subject to the diagonal constraint, with $[{\bolds
{\Sigma}_\bX^{-1}}
|\bolds{\Lambda}]$ as in \eqref{eqn:sigxprior}. This approach is denoted as
\textsc{ebsigmax}. Like \textsc{vanilla}, \textsc{ebsigmax} may only be
implemented when $p\leq\na+\nb$. Finally, let \textsc{ebboth} be the
approach which uses both priors in \eqref{eqn:ridgeprior} and \eqref
{eqn:sigxprior} with \textbf{EWiG} updates for $\lambda$ and $\bm\Lambda
$. These alternatives are all summarized in Table~\ref{tab:2}.

%re2 #&#
\begin{remark} Adaptively estimating the diagonal inverse scale matrix
parameter $\bolds{\Lambda}$ modifies the variance components of $\bX$.
Alternatively, one might apply an \textbf{EWiG} update to the degrees
of freedom parameter, say, $d$, which modifies the partial correlations
of $\bX$. For example, when $d=p+1$, the induced prior on each partial
correlation is uniform on $[-1,1]$ [\citet{gelman.arm}]. Larger values
of $d$ place more prior mass closer to zero. Allowing the data to
specify $d$ is a reasonable alternative; however, we encountered
numerical difficulties in implementing this approach. The \textbf{EWiG}
update cannot be expressed in closed form and must be estimated
numerically. Additionally, the ``complete-data log-likelihood'' in the
$M$-step is often flat, and a wide range of values for $d$ will return
nearly equivalent log-likelihoods.
\end{remark}

%

%s3.3 #&#
\subsection{Estimation under predictive loss}
\label{sec:postsum}
A fitted model may be summarized by measures of uncertainty, for
example, a posterior predictive interval $(\hat Y_\mathrm
{new}^{p_L},\hat Y_\mathrm{new}^{p_H})$, as well as point predictions,
$\hat Y_\mathrm{new}=\beta_0^*+\bX_\mathrm{new}^\top\bbeta^*$ using
summary values $\beta_0^*$ and $\bbeta^*$. These are calculated with
draws from the posterior distribution, $\{\bphi^{(t)}\}$. Predictive
intervals are given by empirical quantiles of $\{Y_\mathrm{new}^{(t)}\}
$, where $\hat
Y_\mathrm{new}^{(t)}=\beta_0^{(t)}+\bX_\mathrm{new}^\top\bbeta
^{(t)} +
\sigma^2{}^{(t)}\varepsilon^{(t)}$ and $\varepsilon^{(t)}\stackrel
{\mathrm{i.i.d.}}{\sim}N\{0,1\}$. For point predictions, a summary value of
$\beta
_0$ is given by $\hat\beta_0=(1/T) \sum_t \beta_0^{(t)}$.\break  For
$\bbeta$,
we minimize posterior predictive loss of $\hat
Y_\mathrm{new}$. Specifically, we define the posterior predictive mean
by $\bbeta^\mathrm{ppm}=\arg\min_{\mathbf{b}}\ev_{\bphi,\bX_\mathrm
{new}|\bU
_\mathrm{obs}} (\bbeta-\break \mathbf{b})^\top\bX_\mathrm{new}\bX_\mathrm
{new}^\top(\bbeta-\mathbf{b})$. This is in contrast\vspace*{1pt} to the posterior mean:
$\bbeta^\mathrm{pm}=\arg\min_{\mathbf{b}}\ev_{\bbeta|\bU_\mathrm
{obs}} (\bbeta
-\mathbf{b})^\top(\bbeta-\mathbf{b})$. Estimates of these quantities are given by
%
%
%
%e11 #&#
%e12 #&#
\begin{eqnarray}
\hat{\bbeta}^\mathrm{ppm}&=& \sum_t \bigl(
\bSigX^{(t)} +\bmuX^{(t)}\bmuX^{(t)}{}^\top
\bigr)^{-1}\sum_t \bigl(
\bSigX^{(t)} +\bmuX^{(t)}\bmuX^{(t)}{}^\top
\bigr)\bbeta^{(t)},\label{eqn:betappm}
\\
\hat{\bbeta}^\mathrm{pm}&=&(1/T)\sum_t
\bbeta^{(t)}\label{eqn:betapm}.
\end{eqnarray}
To summarize, different posterior summaries of $\bbeta$ come from
minimizing different loss functions; we have two estimates of $\bbeta$
for each method and, as a consequence, two choices of point predictions
for $Y_\mathrm{new}$. In contrast, we have only one posterior
predictive interval, that derived from the empirical quantiles of $\{
Y_\mathrm{new}^{(t)}\}$.

%s4 #&#
\section{Simulation study}
\label{sec:simulations}
We conducted a simulation study based upon the motivating data to
evaluate these methods. The assumed model of the data satisfied the
generating model, as given in \eqref{eqn:models}; violations to these
modeling assumptions are considered later. We fixed $n_\mathrm{A}=50$
and $n_\mathrm{B}=400$. The diagonal and off-diagonal elements of
$\bSigX$ were 1 and 0.15, respectively. The regression coefficients
were $\bbeta=\{\frac{j}{100}\}_{j=-49}^{j=49}$ (a diffuse signal) or
$\bbeta=\{\{0.1\}_{k=1}^{k=8},1\}_{j=1}^{j=11}$ (a signal concentrated
in a limited number of coefficients). Values of $R^2$ were either $0.1$
or $0.4$. Given $\bbeta$, $\bSigX$ and $R^2$, $\sigma^2$ was determined
by solving $\bbeta^\top\bSigX\bbeta/(\bbeta^\top\bSigX\bbeta
+\sigma
^2)=R^2$. $\beta_0$ was set to zero. This yielded four unique
simulation settings: two choices each for $\bbeta$ and $R^2$. The
covariates $\bxA$ and $\bxB$ were sampled from $N\{\mathbf{0}_p,\bSigX\}$,
and $\byA|\bxA$ and $\byB|\bxB$ were drawn for each combination of
$\bbeta$ and $\sigma^2$. We set $\psi=0$ and $\nu=1$ and repeated each
of the four settings for $\tau\in(0,2)$, drawing $\bwA|\bxA$ and
$\bwB |\bxB$, the auxiliary data, based on the measurement error model in~\eqref{eqn:models}.

%
%f2 #&#
\begin{figure}

\includegraphics{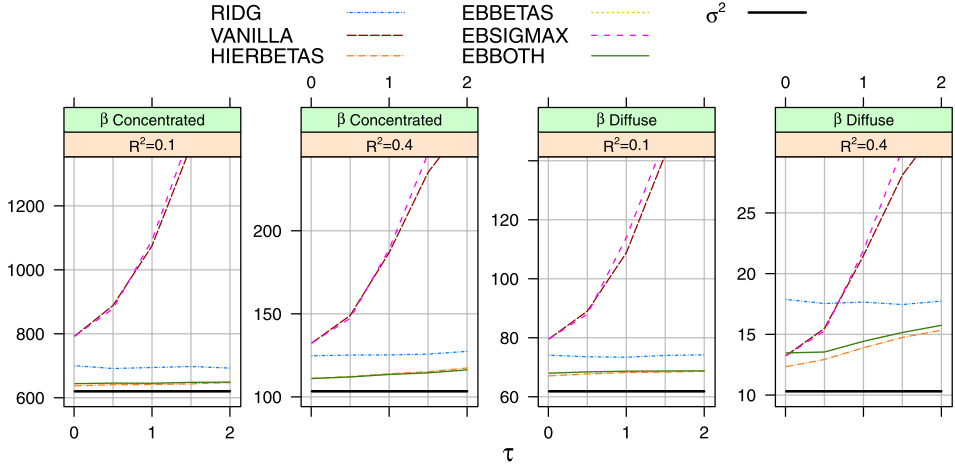}

\caption{$\operatorname{MSPE}(\hat{\bbeta}^\mathrm{ppm})$ plotted
against $\tau$, the standard deviation of the ME model, for four
simulation settings. For each method, $\bbeta$ was estimated from 250
independent training data sets, and MSPE was estimated from 250
validation data sets of size 1000. The thick, solid bar ($\sigma^2$)
corresponds to predictions made using the true generating parameters.
The three best-performing methods, \textsc{hierbetas}, \textsc{ebbetas}
and \textsc{ebboth}, are virtually indistinguishable.}\label{fig:7}
\end{figure}

\begin{figure}

\includegraphics{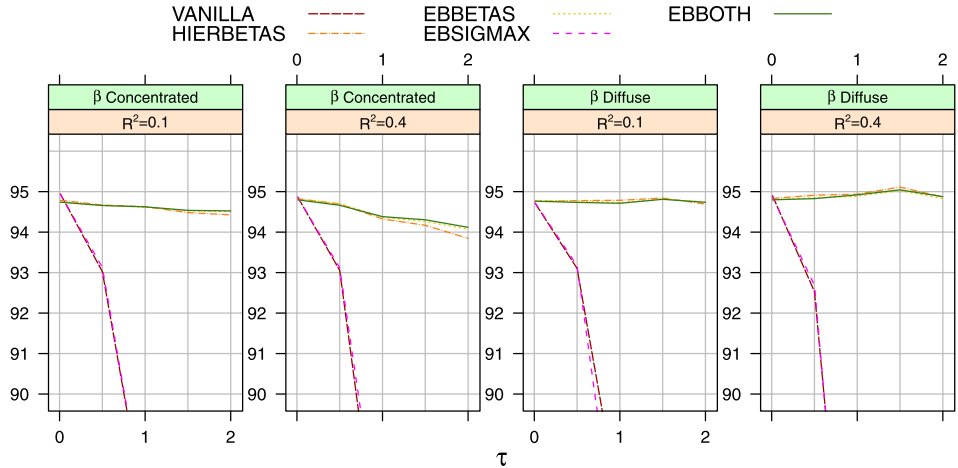}

\caption{Average coverage of prediction intervals
plotted against $\tau$, the standard deviation of the ME model, for
four simulation settings. For each method, prediction intervals were
created using draws of $\bbeta$ from the converged Gibbs sampler, and
coverage was averaged over 250 validation data sets of size 1000.
Nominal coverage is 95. The lines for the three methods that are
closest to maintaining nominal coverage, \textsc{hierbetas}, \textsc
{ebbetas} and \textsc{ebboth}, are virtually indistinguishable.}\label
{fig:coverage}
\end{figure}

After a burn-in period of 2500, we stored 1000 posterior draws. We
calculated $\hat\beta_0$, $\hat{\bbeta}^\mathrm{ppm}$ \eqref{eqn:betappm}
and $\hat{\bbeta}^\mathrm{pm}$ \eqref{eqn:betapm}. For \textsc{vanilla},
\textsc{hierbetas}, \textsc{ebbetas}, \textsc{ebsigmax}, \textsc
{ebboth}, we estimated the MSPE using $\hat{\bbeta}^\mathrm{ppm}$ on 1000
new observations: $\widehat{\operatorname{MSPE}}(\hat\beta_0,\hat
{\bbeta
}^\mathrm
{ppm})=(1/1000)\sum_{j=1}^{1000}(Y_{\mathrm{new},j}-\hat\beta_0-\bX
_{\mathrm
{new},j}^\top\hat{\bbeta}^\mathrm{ppm})^2$. $\{Y_{\mathrm
{new},j},\bX_{\mathrm
{new},j}\}$ are resampled from the same generating distribution for
each simulation. As a comparison, we fit a ridge regression (\textsc
{ridg}) on subsample A only, choosing the tuning parameter with the GCV
function. Figure~\ref{fig:7} plots $\widehat{\operatorname{MSPE}}$,
averaged over
250 simulated data sets, over $\tau$. Smaller values are better, and
the smallest theoretical value is $\sigma^2$, which is also plotted for
reference. We also estimated MSPE using $\hat{\bbeta}^\mathrm{pm}$.
Numerical values are given in Tables S1 and S2, which also contain
results from additional parameter configurations. Finally, we computed
prediction intervals for the new observations (Section~\ref{sec:postsum}). Although frequentist in nature, it is still desirable
for a Bayesian prediction interval to achieve nominal coverage; the
average coverage rates of $Y_{\mathrm{new},j}$, nominally 95\%, are given
in Figure~\ref{fig:coverage}.

From Figure~\ref{fig:7}, \textsc{hierbetas}, \textsc{ebbetas} and
\textsc{ebboth} give about equally good predictions and are
consistently the best overall scenarios. \textsc{ebsigmax}, which
corresponds to shrinkage on ${\bolds{\Sigma}_\bX^{-1}}$ alone,
predicts poorly, and
\textsc{vanilla} does only slightly better. \textsc{ridg} does not beat
the better-performing Bayesian methods. Even though the quality of the
imputations for $\bxB$ depends on the signal in the ME model, the
resulting prediction error of \textsc{hierbetas}, \textsc{ebbetas} and
\textsc{ebboth} varies little over the values of $\tau$ we evaluated.

\emph{Coverage properties}. \textsc{hierbetas}, \textsc{ebbetas}
and \textsc{ebboth} maintain close-to-nominal prediction coverage
(Figure~\ref{fig:coverage}). In contrast, larger values of $\tau$
drastically decrease the coverage of \textsc{vanilla} and \textsc
{ebsigmax}. Prediction intervals for \textsc{ridg} are not automatic
but may be calculated using the bootstrap. This is included in our
primary data analysis.

%
%f3 #&#

\emph{Mean squared error}. The results discussed above and
reported in Figure~\ref{fig:7} use $\hat{\bbeta}^\mathrm{ppm}$, which
minimizes predictive loss, and are evaluated by MSPE. If instead we use
$\hat{\operatorname{MSE}}(\hat{\bbeta}^\mathrm{ppm})$ or
$\hat{\operatorname{MSE}}(\hat{
\bbeta}^\mathrm{pm})$, \textsc{hierbetas}, \textsc{ebbetas} and
\textsc
{ebboth} remain the preferred methods (results not given).

\emph{Computation time}. All Bayesian methods had approximately
equal run-times, each requiring about 110 seconds per data set under
these simulation settings; run-times would increase with $p$, the
dimension of $\bbeta$. While \textsc{ridg} required only 1--2 seconds
for each data set, it does not give automatic prediction intervals, so
a direct comparison of run-times here would be improper. In the data
analysis (Section~\ref{sec:dataanalysis}), we implement a bootstrap
algorithm to construct prediction intervals, allowing for a fair
comparison of computational time. Full computational details are in the
supplemental article [\citet{boonstra.aoassupp}].

\emph{Violations to modeling assumptions.}
As we have noted, these likelihood-based approaches depend on the
assumed model approximately matching the true generating model. We
evaluated robustness by considering the following violations of the
model assumptions: (i) the distribution of $\varepsilon$ is skewed,
shifted to maintain a zero mean: $\varepsilon+1\sim G\{1,1\}$, (ii) the
measurement error model is misspecified $\bW|\bX\sim N_p\{\psi\mathbf{1}_p+\nu
\bX^2,\tau^2\bIp\}$, where we use $\bX^2$ to denote the element-wise
square, or (iii) $\bX$ comes from a mixture of distributions: $\bX
|Z\sim N_p\{ 1_{[Z=2]}(3\times\mathbf{1}_p)-1_{[Z=3]}(3\times\mathbf{1}_p),\bSigX\}
$, where $1_{[\cdot]}$ is the indicator function and $Z\stackrel
{\mathrm{i.i.d.}}{\sim}\operatorname{Unif}\{1,2,3\}$.

The results of these modeling violations are given in Tables S3--S8.
When $\varepsilon$ is skewed (S3, S4), the rankings change little; the
Bayesian ridge methods are equally preferred. The case is similar for
the misspecified measurement error model (S5, S6). When $\bX$ comes from
a mixture of distributions, the results change depending on whether the
signal in $\bbeta$ is concentrated (S7) or diffuse (S8). In the former,
\textsc{ebboth} is best by a large margin for larger values of $\tau$,
even over the other Bayesian ridge methods, \textsc{hierbetas} and
\textsc{ebbetas}. In this case, then, what is required is the joint,
adaptive shrinkage of ${\bolds{\Sigma}_\bX^{-1}}$ and $\bbeta$.
This difference in
performance is not observed when the signal is diffuse (S8), and the
Bayesian ridge methods are all equally good.

A general conclusion of this study is that the shrinkage induced by a
Bayesian ridge regression is adaptable to many scenarios and robust to
modeling violations. The Gibbs sampler allows for the use of the
additional information in subsample~B despite $\bxB$ being missing, and
the ridge prior on $\bbeta$ is effective at controlling variability,
thereby increasing precision in predictions. Most important is that
this holds even when the signals in the outcome model and the ME model
are both very weak, a challenge commonly encountered in the analysis of
genomic data.

%s5 #&#
\section{Data analysis}
\label{sec:dataanalysis}
We now consider the motivating problem of efficiently using the
auxiliary information in the data from \citet{chen.jto}, containing 91
genes representing a broad spectrum of relevant biological functions,
to build a predictive model for survival. Expression using Affymetrix
is measured on 439 tumors, and qRT-PCR measurements are collected on a
subset of 47 of these. Clinical covariates, age, gender and stage of
cancer [I--III], are also available. Because qRT-PCR is the clinically
applicable measurement for future observations, the goal is\vadjust{\goodbreak} a qRT-PCR${}+{}$clinical
covariate model for predicting survival time after surgery. An
independent cohort of 101 tumors with qRT-PCR measurements and clinical
covariates is available for validation. After some necessary
preprocessing of the data, as described in the supplemental article
[\citet{boonstra.aoassupp}], the available data had $\na=47$, $\nb=389$,
and the validation sample is size 100.

%
%t3 #&#
\begin{table}[b]\vspace*{-3pt}
\caption{Results from lung adenocarcinoma analysis.
$\widehat{\operatorname{MSPE}}$ is the empirical prediction error in the validation
data, SIBS is the Scaled Integrated Brier Score, Avg. Coverage is
average coverage of the prediction intervals, $\operatorname{Avg}(\hat Y_\mathrm
{new}^{97.5}-\hat Y_\mathrm{new}^{2.5})$ gives the average prediction
interval length for the validation sample, and Computation gives the
time, in seconds, to calculate coefficient estimates and prediction intervals}
\label{tab:6}
\begin{tabular*}{\textwidth}{@{\extracolsep{\fill}}lcccccc@{}}
\hline
&\textbf{\textsc{ridg}}&\textbf{\textsc{vanilla}}&\textbf{\textsc{hierbetas}}&\textbf{\textsc
{ebbetas}}&\textbf{\textsc{ebsigmax}}&\textbf{\textsc{ebboth}}\\
\hline
$\widehat{\operatorname{MSPE}}(\hat\beta_0,\hat{\bbeta}^\mathrm{ppm})$&
0.620& 1.251 &
0.555 & 0.555 & 1.230 & 0.561 \\
$\widehat{\operatorname{MSPE}}(\hat\beta_0,\hat{\bbeta}^\mathrm{pm})$&
--&1.768& 0.559&
0.558 & 1.932 &0.560 \\
$\operatorname{SIBS}(\hat{\bbeta}^\mathrm{ppm})$ & 0.544 & 0.629 & 0.394 & 0.393 &
0.632 &
0.396\\
$\operatorname{SIBS}(\hat{\bbeta}^\mathrm{pm})$ & -- & 0.796 & 0.395 & 0.395 &
0.848 &
0.395\\
Avg. Coverage &0.92& 0.88 &0.96& 0.97& 0.87& 0.96\\
$\operatorname{Avg}(\hat Y_\mathrm{new}^{97.5}-\hat Y_\mathrm{new}^{2.5})$&3.37& 3.98
& 3.11 & 3.11 & 3.93 & 3.09 \\
Computation (sec) &298 &268 &269 & 268&269 &269\\
\hline
\end{tabular*}
\end{table}

Because our methodology was developed for continuous outcomes,
censoring necessitated some adjustments to the data in order to fit our
models. We first imputed each censored log-survival time from a linear
model of the clinical covariates, conditional upon the censoring time.
This model was fit to the training data, but censored survival times in
both the training and validation data were imputed. Given completed
log-survival times, we refit this same model and calculated residuals
from both the training and validation data. These residuals were
considered as outcomes, and the question is whether any additional
variation in the residuals is explained by gene expression. While there
are other ways of dealing with coarsened data and additional covariates
in the likelihood-based framework, processing the data this way allows
for \textsc{ridg} to serve as a reference. To more realistically model
the data, we allow for a gene-specific ME model: $w_{ij} = \psi_j +
\nu
_j x_{ij} +\tau\xi_{ij}$. To incorporate this modification into our
model, we put independent flat priors on $\psi_j$ and $\nu_j$,
$j=1,\ldots,p$. The modified Gibbs steps are included in the
supplemental article [\citet{boonstra.aoassupp}].

We applied each Bayesian approach, running each chain of the Gibbs
sampler for 4000 iterations and storing posterior draws from the
subsequent 4000 iterations. Table~\ref{tab:6} presents numerical
results: the estimated MSPE from predicting the uncensored residuals in
the validation data and the average prediction coverage of these
residuals. Additionally, Table~\ref{tab:6} presents the Scaled
Integrated Brier Score [SIBS, \citet{graf.statmed}], which is a scoring
method for right-censored data, on the original, unadjusted validation
data.\vadjust{\goodbreak}

To calculate the SIBS, which is a function of predicted survival
probabilities, we used the survival function from the Normal
distribution, estimating the mean log-survival time by adding the
linear predictor of the genomic data to the linear predictor of the
clinical covariates. At each unique time of last follow-up (either time
of death or censoring), the squared difference in predicted survival
probability for each individual minus current dead/alive status was
calculated and averaged over all individuals and integrated over all
time points, with censored individuals only contributing to the
calculation of the score until their censoring time. This quantity was
scaled by a reference score, that from plugging in 0.5 as a predicted
survival probability everywhere, to get the SIBS. Thus, any model that
does better than random guessing has a SIBS in the interval (0,1), and
a smaller SIBS is better.

Based upon MSPE, \textsc{hierbetas}, \textsc{ebbetas} and \textsc
{ebboth} were about equally good, with MSPEs of 0.555, 0.555 and 0.561,
respectively, using $\hat{\bbeta}^\mathrm{ppm}$. These MSPEs are smaller
than those from \textsc{ridg} (0.620) as well as \textsc{vanilla}
(1.251), and \textsc{ebsigmax} (1.230). Using $\hat{\bbeta}^\mathrm{pm}$,
the estimated posterior mean of $\bbeta$, the three best methods gave
almost identical results, while \textsc{vanilla} and \textsc{ebsigmax}
had worse prediction error. Similarly, \textsc{hierbetas}, \textsc
{ebbetas} and \textsc{ebboth} had the smallest SIBS (resp.,
0.394, 0.393 and 0.396), and the remaining methods had larger SIBS.

Considering coverage of the prediction intervals, \textsc{hierbetas}
(0.96), \textsc{ebbetas} (0.97) and \textsc{ebboth} (0.96) all had
rates close to their nominal values, and their prediction intervals
widths are smallest. This contrasts with \textsc{vanilla} and \textsc
{ebsigmax}, whose coverage rates are less than nominal (0.88, 0.87). We
created prediction intervals for \textsc{ridg} using a bootstrap
algorithm; the resulting coverage is 0.92. The required computational
time is 298 seconds for \textsc{ridg}, including the bootstrap
algorithm to calculate prediction intervals, and about 268--269 seconds
for each Bayesian method. Although $p$, $\na$ and $\nb$ were about the
same as in the simulation study, fitting the methods took longer (268
vs. 110 seconds) because the number of total MCMC iterations increased
(8000 vs. 3500).

%f4 #&#
\begin{figure}

\includegraphics{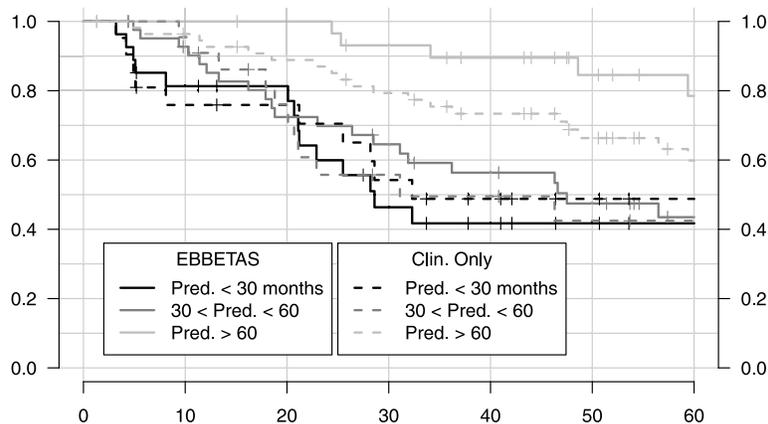}

\caption{Comparison of risk-indexed Kaplan--Meier plots.
For both \textsc{ebbetas} and an accelerated failure time model using
only the clinical covariates, the validation data was grouped based on
predicted survival time (less than 30 months, between 30 and 60 months,
and longer than 60 months).}\label{fig:3}
\end{figure}

To summarize the analysis thus far, a Bayesian ridge regression, which
uses all observations in the data, offers better overall predictive
performance in our validation data and, compared to a ridge regression
on the complete observations alone, narrower prediction intervals that
still achieve nominal coverage. This is a reflection of the extra
information that is available in the incomplete observations. Beyond
the question of \emph{how} to use the auxiliary genomic information in
a prediction model, which has been already been covered in detail, more
fundamental to the application is whether one of the Bayesian ridge
regressions, for example, \textsc{ebbetas}, can do better than an
analysis using clinical covariates alone, of which complete information
is available on all observations. The natural comparison would be an
accelerated failure time (AFT) regression, modeling censored
log-survival time as a linear function of the clinical covariates and
gaussian noise. Predictions from this AFT model could be directly
compared to the outcome model in~\eqref{eqn:linearmodel}.

The SIBS from fitting the AFT model is 0.394, nearly equal to that of
\textsc{ebbetas}. Exploring this comparison further, Figure~\ref{fig:3}
gives risk-indexed Kaplan--Meier plots of the validation data, comparing
predictions using \textsc{ebbetas} (calculated by adding together the
genomic linear predictors to the clinical covariate linear predictors
described at the beginning of this section) to that of the AFT model.
For each model, patients in the validation sample were indexed based on
the their predicted survival time: less than 30 months, between 30 and
60 months, or longer than 60 months. From the figure, the clearest
distinction is in the low-risk group, those predicted to live longer
than 60 months. In the low-risk, ``$>$60 month'' group as defined by
\textsc{ebbetas}, 25 out of 31 patients, or about 80\%, were alive at
60 months' time. This contrasts with the AFT model: 56 patients were
predicted to live beyond 60 months, and 36, or about 64\%, were alive
at 60 months' time. Also distinctive is that the survival curves for
the medium- and high-risk groups of the AFT model cross several times
and generally show less separation compared to \textsc{ebbetas}. The
estimated median survival times for these two groups are 28.6 (high)
and 47.5 (med.) months under the \textsc{ebbetas}-based grouping versus
32.3 (high) and 31.1 (med.) under the AFT grouping. Thus, despite
nearly equal values of the SIBS, which are aggregate measures of
predictive performance, \textsc{ebbetas} appears to have better
individual predictions and discrimination between the three groups.

%

%s6 #&#
\section{Discussion}
Driven by a need to incorporate genomic information into prediction
models, we have considered the problem of shrinkage in a model with
many covariates when a large proportion of the data are missing.
Predictions for future observations are of primary interest. We discuss
the primary contributions of this paper in two parts as follows.

%s6.1 #&#
\subsection{Shrinkage via the Gibbs sampler}
A likelihood-based approach confers a number of advantages, these being
the inclusion of shrinkage into the likelihood and the proper
accounting of uncertainty in predictions coming from the unobserved
data. A number of existing Bayesian approaches for the treatment of
missing data and/or implementation of shrinkage methods are easily
adapted here. We have shown how two such approaches, the Monte Carlo EM
[\citet{wei.jasa}], a Gibbs sampler which multiply imputes missing data,
and the Empirical Bayes Gibbs Sampler [\citet{casella.biostatistics}], a
Gibbs sampler which adaptively shrinks parameter estimates, generalize
to the same algorithm, which we call EM-within-Gibbs.

%
%f5 #&#
\begin{figure}

\includegraphics{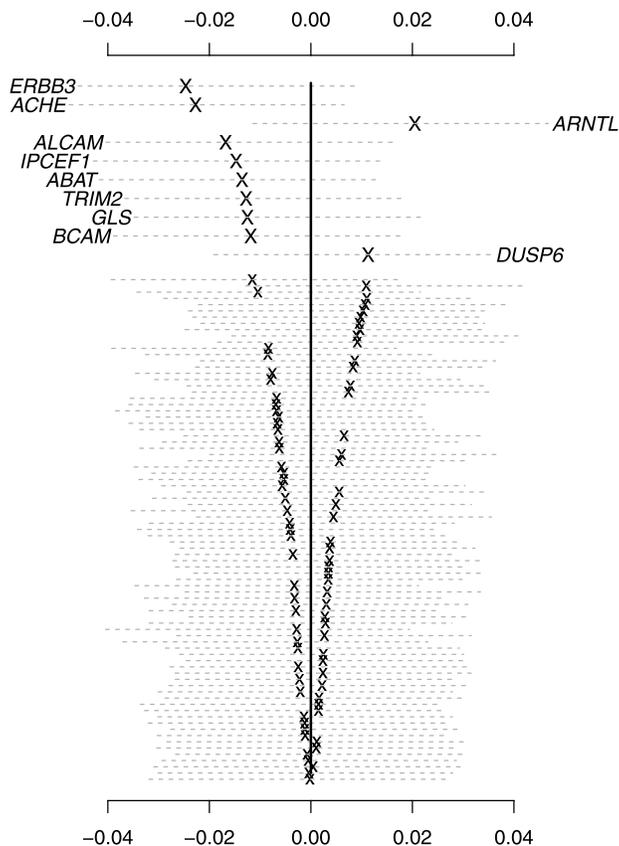}

\caption{Coefficient estimates (X) and 95\%
credible intervals ($-$ $-$) of the 91 genes according to \textsc{ebbetas},
ordered from top to bottom by the magnitude of the ratio of posterior
predictive mean to posterior standard deviation. The top ten genes are
highlighted and annotated.}\label{fig:geneontology}
\end{figure}

We proposed specific choices of prior specification aimed at improving
prediction with shrinkage methods. The various flavors of the Bayesian
ridge, denoted as \textsc{hierbetas}, \textsc{ebbetas} and \textsc
{ebboth}, stand out as the methods of choice, indicating that shrinkage
of $\bbeta$, which is the vector of regression coefficients in the
outcome model, is most crucial, over and above no shrinkage at all
(\textsc{vanilla}) or shrinkage of ${\bolds{\Sigma}_\bX^{-1}}$
alone (\textsc
{ebsigmax}). Our simulation study and data analysis showed the Bayesian
ridge to be best under a number of scenarios using several criteria,
including MSPE and prediction coverage, and robust to several modeling
violations. In addition, the Bayesian ridge does not require $p\leq\na
+\nb$, in contrast to \textsc{vanilla} or \textsc{ebsigmax}. As for the
specific choice of which Bayesian ridge regression is best, we found
little evidence to recommend any one variant.

That shrinkage of ${\bolds{\Sigma}_\bX^{-1}}$ alone, as we have
implemented it, does
not improve predictions (and sometimes actually worsens predictions)
may be due to the specific nature of the shrinkage we implemented. The
mean of the conditional distribution of ${\bolds{\Sigma}_\bX^{-1}}$
given in \eqref
{eqn:sigxconddist} is a convex combination of $\bLambda/(3p)$, which is
the inverse of its prior mean, and the sample variance of $\bxA$ and
$\bxB$. In contrast, ridge regression may be viewed as simply adding
$\lambda\bIp$ to the sample variance of the covariates. The Wishart
prior cannot mimic this effect, and the construction of a different,
nonconjugate prior for ${\bolds{\Sigma}_\bX^{-1}}$ may be required
to induce ridge-type
shrinkage.

%s6.2 #&#
\subsection{Using genomic information in prediction models}
\label{sec:discussion2}
Figure~\ref{fig:geneontology} plots coefficient estimates and 95\%
credible intervals for the 91 genes according to \textsc{ebbetas}. They
are ordered by the ratio of their posterior mean to posterior standard
deviation, an estimate of statistical significance. The ten most
significant genes are annotated, according to the \texttt{R} package \texttt{annotate}
[\citet{annotate.cran}]. Even the most significant gene,
\textit{ERBB3}, is not significant at the 0.05 level. Although these
are preselected genes that were deliberately chosen to represent a wide
spectrum\vadjust{\goodbreak} of biological functions, many of which have already been
implicated in different cancers, this lack of significance for
individual genes is not unexpected. The genomic effect is likely to be
at the pathway-level rather than individual expressions, which a plot
like Figure~\ref{fig:geneontology} is too coarse to detect. Despite
this lack of individual significance, the small genomic effects
collectively yield an overall improvement, albeit small, in predictive
ability when the information is properly incorporated, and the Bayesian
ridge regression appears best-equipped to do so.


\begin{supplement}[id=suppA]
\stitle{Supplemental article}
\slink[doi]{10.1214/13-AOAS668SUPP} %[doi,text={...}] - jei reikia
%suskaldyti doi
\sdatatype{.pdf}
\sfilename{aoas668\_supp.pdf}
\sdescription{Here we give the full derivation of the Gibbs steps,
computational details and the results from the simulation study. The
data from Section~\ref{sec:dataanalysis} and the code for its analysis
are available at \url{http://www-personal.umich.edu/\textasciitilde philb}.}
\end{supplement}

%
% imsref loaded by akundreckaite, 2013-09-03 15:32:36
%

%

\printaddresses

\end{document}